\documentstyle[11pt,paspconf,epsfig]{article}

\def\kmsmpc{\,{\rm km\,s^{-1}\,Mpc^{-1}}}

\def\msun{\,{\rm M_\odot}}

\def\sfrd{\,{\rm M_\odot\,yr^{-1}\,Mpc^{-3}}}
\def\rd{\,{\rm yr^{-1}\,Mpc^{-3}}}

\def\etal{{et al.\ }}

\def\spose#1{\hbox to 0pt{#1\hss}}
\def\lta{\mathrel{\spose{\lower 3pt\hbox{$\mathchar"218$}}
     \raise 2.0pt\hbox{$\mathchar"13C$}}}
\def\gta{\mathrel{\spose{\lower 3pt\hbox{$\mathchar"218$}}
     \raise 2.0pt\hbox{$\mathchar"13E$}}}
 
\begin{document}

\title{Galaxy Evolution and the Cosmic Rate of Supernovae}
\author{Piero Madau}
\affil{Space Telescope Science Institute, 3700 San Martin Drive, Baltimore,
MD 21218}

\begin{abstract}

Ongoing searches for supernovae (SNe) at cosmological distances have recently
started to provide a link between SN Ia statistics and galaxy evolution. In 
this talk I will use recent estimates of the global history of star formation 
to compute the
theoretical Type Ia and Type II SN rates as a function of cosmic time.
I will compare the predicted values  with the
rates observed in the range $0\le z\le0.4$, and show how accurate
measurements of the frequency of SN events at intermediate redshifts are 
valuable probes of the evolution of the stellar birthrate in the universe
and the nature of Type Ia progenitors. The {\it Next Generation Space 
Telescope} should detect of order 20 Type II SNe per field per year in the 
interval $1<z<4$.
 
\end{abstract}


\section{Introduction}

The evolution with redshift of the rate of supernovae (SNe) contains unique 
information on the star formation history
of the universe, the initial mass function (IMF) of stars, and the nature of
the binary companion in Type Ia events. All are essential ingredients for
understanding galaxy formation, cosmic chemical evolution, and the mechanisms
which determined the efficiency of the conversion of gas into stars in galaxies
at various epochs (Madau \etal 1996; Madau \etal 1997a; A. Renzini, this 
volume). While the frequency of ``core-collapse supernovae'', SN~II and
possibly SN~Ib/c, which have short-lived progenitors, is essentially related, 
for a
given IMF, to the instantaneous stellar birthrate of massive stars, Type Ia SNe
-- which are believed to result from the thermonuclear disruption of C-O white
dwarfs in binary systems -- follow a slower evolutionary clock, and can then
be used as a probe of the past history of star formation in galaxies 
(Ruiz-Lapuente \etal 1997; Yungelson \& Livio 1997).
 
The tremendous progress in our understanding of faint galaxy data made possible
by the combination of {\it HST} deep imaging (Williams \etal 1996) and
ground-based spectroscopy (Lilly \etal 1996; Ellis \etal 1996; Cowie \etal
1996; Steidel \etal 1996), together with the recent detection of Type Ia SNe at
cosmological distances (Garnavich \etal 1997; Perlmutter \etal
1997), allow for the first time a detailed comparison between the SN rates
self-consistently predicted by stellar evolution models that reproduce the
optical spectrophotometric properties of field galaxies  and the observed 
values (Madau \etal 1997b). In this talk I will show how
accurate measurements of the frequencies of Type II(+Ib/c) and Ia SNe 
at low and intermediate redshifts could be used as an independent test for 
the star
formation and heavy element enrichment history of the universe,  and
significantly improve our understanding of the intrinsic nature and age
of the populations involved in the SN explosions. At $z>2$, the
biggest uncertainty in our estimates of the star formation density
is probably associated with dust extinction. A determination of the amount
of star formation at early epochs is of crucial importance, as the two
competing scenarios
for galaxy formation, ``monolithic collapse'' -- where spheroidal systems formed
early and rapidly, experiencing a bright starburst phase at high-$z$ (Eggen,
Lynden-Bell, \& Sandage 1962) -- and ``hierarchical clustering'' -- where ellipticals form continuosly by the merger of disk/bulge
systems (White \& Frenk 1991) and most galaxies never
experience star formation rates in excess of a few solar masses per year (Baugh
\etal 1997) -- make very different predictions in this regard. I will show how,
by detecting Type II SNe at high-$z$, the {\it Next Generation Space 
Telescope} should provide a unique test for distinguishing between different 
scenarios of galaxy formation.

\begin{figure}[b!]
\centerline{\epsfig{file=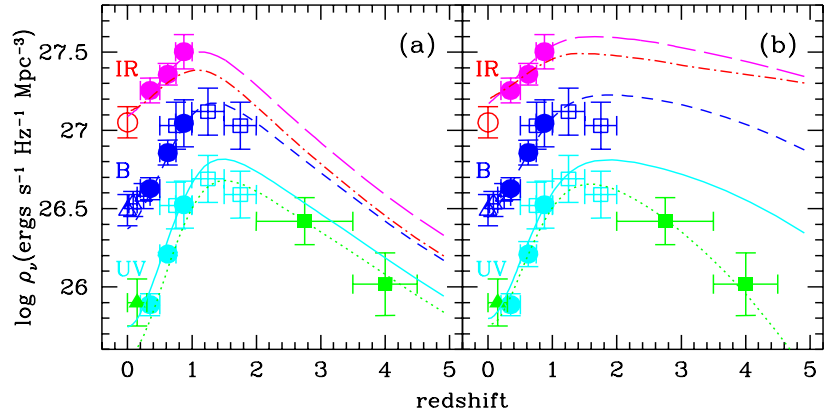,height=4.3in,width=4.5in}}
\vspace{-4.3cm}
\caption{Evolution of the observed comoving luminosity density at rest-frame 
wavelengths of 0.15 ({\it dotted line}), 0.28 ({\it solid line}), 0.44 
({\it short-dashed line}), 1.0 ({\it long-dashed line}), and 2.2 ({\it 
dot-dashed line}) \micron. The data points with error bars are taken from 
Lilly \etal (1996) ({\it filled dots}), Connolly \etal (1997) ({\it empty 
squares}), Madau \etal (1996) ({\it filled squares}), Ellis 
\etal (1996) ({\it empty triangles}), Treyer \etal (1997) ({\it filled 
triangle}), and Gardner \etal (1997) ({\it empty dot}). A flat cosmology with
$q_0=0.5$ and $H_0=50\,h_{50}\kmsmpc$ was adopted. ({\it a}) The model 
assumes a Salpeter IMF, SMC-type dust in a foreground screen, and a 
universal $E(B-V)=0.1$. ({\it b}) The model -- designed to mimick a
monolithic collapse scenario -- assumes a Salpeter IMF and a dust opacity
which increases rapidly with redshift, $E(B-V)=0.011(1+z)^{2.2}$. 
\label{fig1}}
\end{figure}

\section{Cosmic Star Formation History}

I will model the emission
history of field galaxies at ultraviolet, optical, and near-infrared
wavelengths by tracing the evolution with cosmic time of their luminosity
density, 
\begin{equation}
\rho_\nu(z)=\int_0^\infty L_\nu y(L_\nu,z)dL_\nu=\Gamma(2+\alpha)y_*L_*,
\end{equation}
where $y(L_\nu,z)$ is the best-fit Schechter luminosity function in each
redshift bin. The integrated light radiated per unit volume from the entire
galaxy population is an average over cosmic time of the stochastic, possibly
short-lived star formation episodes of individual galaxies, and follows a
relatively simple dependence on redshift. Madau \etal (1997a) have shown how 
a stellar evolution model, defined by a time-dependent star formation rate per 
unit volume, $\psi(t)$, a universal IMF, $\phi(m)$, and some amount of dust 
reddening, can actually reproduce the optical data reasonably well. In such a
system, the luminosity density at time $t$ is given by the convolution integral
\begin{equation} 
\rho_\nu(t)=p_{\rm esc}\int^t_0 l_\nu(t')\psi(t-t')dt', 
\end{equation} 
where $l_\nu(t')$ is the specific luminosity radiated per unit initial mass
by a generation of stars with age $t'$, and $p_{\rm esc}$ is a
time-independent term equal to the fraction of emitted photons which are not
absorbed by dust. 
The function $\psi(z)$ is derived from the observed UV luminosity density, and
is then used as input to the population synthesis code of Bruzual \& Charlot 
(1997). 

Figure 1{\it a} shows the model predictions for the evolution of $\rho_\nu$ for
a Salpeter function, $E(B-V)=0.1$ with SMC-type dust (in this case, the
observed UV luminosities must be corrected upwards by a factor of 1.4 at 2800
\AA\, and 2.1 at 1500 \AA), and a star formation history which traces the rise,
peak, and sharp drop of the UV emissivity.\footnote{Although in our calculations
the IMF extends down to $m_l=0.1 \msun$, stars below 0.8 $\msun$ make only a
small contribution to the emitted light. This introduces a
significant uncertainty in our estimates of the total stellar birthrate. For
example, in the case of a Salpeter IMF with $m_l=0.5 \msun$, the inferred star
formation rate, $\psi\propto m_l^{-0.35}$, would decrease by a factor of
1.86.}~ For simplicity, the metallicity was fixed to solar values and the IMF truncated at 0.1
and 125 $\msun$. The data points show the observed luminosity density in six
broad passbands centered around 0.15, 0.20, 0.28, 0.44, 1.0, and 2.2 \micron. 
The model is able to account for the entire background light recorded in 
the galaxy counts down to the very faint magnitude levels probed by the 
{\it Hubble Deep Field} (HDF), and produces visible mass-to-light ratios at the present epoch
which are consistent with the values observed in nearby galaxies of various
morphological types. The bulk
($\gta 60\%$ by mass) of the stars present today formed relatively recently ($z\lta
1.5$), consistently with the expectations from a broad class of hierarchical
clustering cosmologies (Baugh \etal 1997), and in good agreement with the low
level of metal enrichment observed at high redshifts in the damped
Lyman-$\alpha$ systems (Pettini \etal 1997a). 

One of the biggest uncertainties in our understanding of the evolution of
luminous matter in the universe is represented by the poorly constrained
amount of starlight that was absorbed by dust and reradiated in the far-IR
at early epochs. While the model of Figure 1{\it a} reproduces
quite well the rest-frame UV colors of high-$z$ objects in the HDF, the 
prescription for a ``correct'' de-reddening of the Lyman-break galaxies 
is the subject of an ongoing debate (Pettini \etal 1997b and references
therein). Figure 1{\it b} shows the model predictions for a monolithic collapse 
scenario, where half of the present-day stars were formed at $z>2.5$ and
were enshrouded by dust. Consistency with the HDF ``dropout''
analysis has been obtained
assuming a dust extinction which increases rapidly with redshift, 
$E(B-V)=0.011(1+z)^{2.2}$. This results in a correction to the rate of 
star formation by a factor of $\approx 5$ at $z=3$ and $\approx 15$ at $z=4$.
While still consistent with the global history of light, this model appears
to overpredict the metal mass density at high redshifts as sampled by QSO
absorbers. Ultimately, it should be possible to set some constraints on the 
total amount of star 
formation that is hidden by dust over the entire history of the universe by
looking at the cosmic infrared background (G. De Zotti and B. Guiderdoni, 
this volume).

In the next section we shall compute the expected evolution with cosmic time of
the Type Ia and II/Ib,c supernova frequencies for the two star formation
histories discussed above. By focusing on the integrated
light radiated by the galaxy population as a whole, our approach will not
specifically address the evolution and the SN rates of particular subclasses of
objects, like the oldest ellipticals or low-surface brightness galaxies, whose
star formation history may have differed significantly from the global average.

\section{Type Ia and II(+Ib/c) Supernova Rates}

Single stars with mass $>8\msun$ evolve rapidly ($\lta 50\,$ Myr) through all
phases of central nuclear burning, ending their life as Type II SNe with
different characteristics depending on the progenitor mass. For a Salpeter IMF
the core-collapse
supernova rate can be related to the stellar birthrate according to 
\begin{equation}
{\rm SNR}_{\rm II}(t)=\psi(t){\int_8^{125} dm \phi(m)\over \int_{0.1}^{125} 
dm m \phi(m)}=0.0074\times \left[{\psi(t)\over \sfrd} \right]\, \rd.
\end{equation} 
It is worth noting at this stage that our model predictions for the 
frequency of Type II events are largely independent of the assumed IMF. This
follows from the fact that the rest-frame UV continuum emission -- which is
used as an indicator of the instantaneous star formation rate -- from
all but the oldest galaxies is entirely dominated by massive stars on
the main sequence, the same stars which later give origin to a Type II SN. 

The specific evolutionary history leading to a Type Ia event remains instead 
an unsettled question. SN Ia are believed to result from the explosion of C-O
white dwarfs (WDs) triggered by the accretion of material from a companion, the
nature of which is still unknown (see Ruiz-Lapuente \etal 1997 for a recent
review). In a {\it double degenerate} (DD) system, for example, such elusive
companion is another WD: the exploding WD reaches the Chandrasekhar limit and
carbon ignition occurs at its center (Iben \& Tutukov 1984). In the {\it single
degenerate} (SD) model instead, the companion is a nondegenerate, evolved star
that fills its Roche lobe and pours hydrogen or helium onto the WD (Whelan \&
Iben 1973; Iben \& Tutukov 1984). While in the latter the clock for the
explosion is set by the lifetime of the primary star, and, e.g., by how long it
takes to the companion to evolve and fill its Roche lobe, in the former it is
controlled by the lifetime of the primary star and by the time it takes to shorten the separation of the two WDs as a
result of gravitational wave radiation. The evolution of the rate depends then,
among other things, on the unknown mass distribution of the secondary binary
components in the SD model, or on the distribution of the initial separations
of the two WDs in the DD model. 

\begin{figure}
\centerline{\epsfig{file=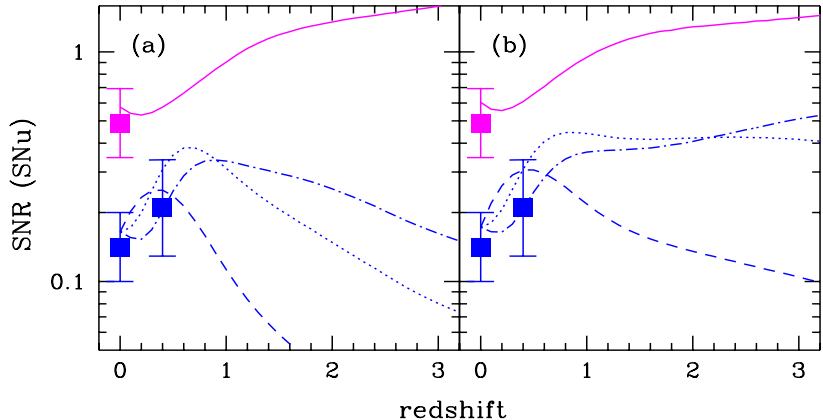,height=4.3in,width=4.5in}}
\vspace{-4.3cm}
\caption{Predicted Type Ia and II(+Ib/c) rest-frame frequencies as a function 
of redshift (from Madau \etal 1997b). The rates are 
normalized to the {\it emitted} blue luminosity 
density. {\it Solid line}: SN II rate.	{\it Dashed-dotted line}: SN Ia 
rate with $\tau=0.3$ Gyr. {\it Dotted line}: SN Ia rate with $\tau=1$ Gyr. {\it Dashed line}: SN Ia rate with $\tau=3$ Gyr. The data points 
with error bars have been derived from the measurements of Cappellaro 
\etal (1997), Tammann \etal (1994), Evans \etal (1989), and Pain \etal (1997), and have been weighted according to 
the local blue luminosity function by spectral type of Heyl \etal (1997). ({\it a}) Model predictions for the merging scenario of Figure 1{\it a}.  ({\it b}) 
Same for the monolithic collapse scenario of Figure 1{\it b}.
\label{fig2}}
\end{figure}

To shed light into the identification issue and, in particular, on the 
clock-mechanism for the explosion of Type Ia's, a more empirical approach will
be adopted here. I will parametrize the rate of Type Ia's in terms of 
a characteristic explosion timescale, $\tau$ -- which defines an explosion
probability per WD assumed to be {\it independent} of time -- and an explosion
efficiency, $\eta$. The former accounts for the time it takes in the various
models to go from a {\it newly born} (primary) WD to the SN explosion itself: a
spread of ``delay'' times results from the combination of a variety of initial
conditions, such as the mass ratio of the binary system, the distribution of
initial separations, the influence of metallicity on the mass transfer rate and
accretion efficiency, etc. The latter simply accounts for the fraction of stars 
in binary systems that, because of unfavorable
initial conditions, will never undergo a SN Ia explosion. We consider as
possible progenitors all systems in which the primary star has an {\it initial}
mass higher than $m_{\rm min}=3\,M_\odot$ (final mass $\ge 0.72\,M_\odot$,
Weidemann 1987) and lower than $m_{\rm max}=8\,M_\odot$: stars less massive
than $3\,M_\odot$ will not produce a catastrophic event even if the companion
has comparable mass, while stars more massive than $8\,M_\odot$ will undergo
core collapse, generating a Type II explosion. 

With these assumptions the rate of Type Ia events at any one time will be
given by the sum of the explosions of all the binary WDs produced in the
past that have not had the time to explode yet, i.e. 
\begin{equation}
{\rm SNR}_{\rm Ia}(t)={\eta \int^t_0 \psi(t')dt'\int_{m_c} 
^{m_{\rm max}} \exp(-{t-t'-t_m\over \tau})\phi(m)dm 
\over \tau \int\phi(m)dm},
\end{equation}
where $m_c\equiv{\rm max}[m_{\rm min},m(t')]$, $m(t')=(10\,{\rm Gyr}/t')^{0.4}$ 
is the minimum mass of a star that 
reaches the WD phase at time $t'$, and $t_m=10\,{\rm Gyr}/m^{2.5}$ is the 
standard lifetime of a star of mass $m$ (all stellar masses are expressed 
in solar units). For a fixed initial mass $m$, the frequency of Type Ia events
peaks at an epoch that reflect an ``effective'' delay $\Delta t\approx \tau+t_m$
from stellar birth. A prompter (smaller $\tau$) explosion results in a 
higher SN Ia rate at early epochs. 

\section{Comparison with the Data}

Observed rates of SNe are normally given in units of SNu, one SNu corresponding
to 1 SN per 100 years per $10^{10} L_{B\odot}$. Since a galaxy luminosity
depends on $H_0$, there is a factor $h_{50}^2$ involved in the inferred SN
frequencies. 
In Figure 2 we show the predicted Type Ia and II(+Ib/c) rest-frame
frequencies as a function of redshift. Expressed in SNu, the Type II rate
is basically proportional to the ratio between the UV and blue galaxy
luminosity densities, and is therefore independent of cosmology. Unlike the SN
frequency per unit volume, which will trace the evolution of the stellar 
birthrate, the  frequency of Type II events per unit blue luminosity
is a monotonic increasing function of redshift, and depends only weakly on the 
assumed star formation history. The Type Ia rates plotted in the figure
assume characteristic ``delay'' timescales after the collapse of the primary 
star to a WD equal to $\tau=0.3, 1$ and 3 Gyr, which virtually encompass 
all relevant possibilities. The SN Ia explosion efficiency was left as an
adjustable parameter to reproduce the observed ratio of SN II to SN Ia
explosion rates in the local Universe, ${\rm SNR}_{\rm II}/{\rm SNR}_{\rm
Ia}\approx 3.5$, $5\%<\eta<10\%$ for the adopted models. Note how,
relative to the merging scenario, the monolithic collapse model predicts SN Ia
rates (in SNU) that are, in the $\tau=0.3\,$ Gyr case, a factor of 1.6 and 4.9
higher at $z=2$ and 4, respectively. 

\section{Conclusions}

We have investigated the link between SN statistics and galaxy evolution. 
Using recent estimates of the star formation history of field galaxies
and a simple model for the evolutionary history of the binary system 
leading to a Type Ia event, I have computed the
theoretical Type Ia and Type II SN rates as a function of cosmic time. Our 
main results can be summarized as follows.

\begin{itemize}

\item At the present epoch, the predicted Type II(+Ib/c) frequency 
appears to match quite well, to within the errors, the observed local value. 
The SN II rate is a sensitive function of the lower mass cutoff of the 
progenitors, $m_l$. Values as low as $m_l=6\,\msun$ (Chiosi \etal 1992) or 
as high as  $m_l=11\,\msun$ (Nomoto 1984) have been proposed 
in the literature: adopting a lower  mass limit of 6 or 11 $\msun$ would
increase or reduce our Type II rates by a factor 1.5, respectively.

\item At low and intermediate redshifts, a comparison between the predicted
and observed rates shows that the best match is obtained for a class of Type Ia
progenitors in which the explosion epoch is delayed by a characteristic
timescale $\sim 0.5$ Gyr after the collapse of the primary stars to a WD, and
a SN Ia efficiency of order $5-10\%$. Accurate measurements of SN rates in the
range $0.5\lta z\lta 1.5$ would dramatically improve our understanding of the
nature and physics of SN~Ia explosions. Ongoing searches and studies of 
distant SNe (Garnavich \etal 1997; Perlmutter \etal 1997)
should provide these rates in the next few years.

\begin{figure}
\centerline{\epsfig{file=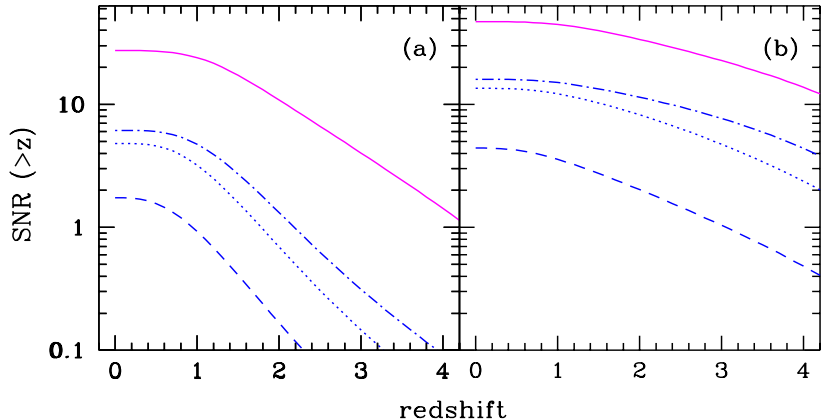,height=4.3in,width=4.5in}}
\vspace{-4.3cm}
\caption{Predicted cumulative number of Type Ia and II(+Ib/c) SNe above a
given redshift $z$ in a $4'\times 4'$ field (from Madau \etal 1997b). 
{\it Solid line}: Type II's.
{\it Dashed-dotted line}: Type Ia's with $\tau=0.3$ Gyr. {\it Dotted 
line}: Type Ia's with $\tau=1$ Gyr. {\it Dashed line}: Type Ia's with 
$\tau=3$ Gyr.  The effect of dust extinction on the detectability of SNe has
not been included in the models. ({\it a}) Model predictions for the 
merging scenario of Figure 1{\it a}.  ({\it b}) Same for the monolithic 
collapse scenario of Figure 1{\it b}.
\label{fig3}}
\end{figure}

\item At higher redshifts, $2<z<4$, the detection of Type II SNe must await the
{\it Next Generation Space Telescope} (NGST). A SN II has a typical peak
magnitude $M(B)\approx -17$ (e.g. Patat \etal 1994): placed at $z=3$, such an
event would give rise to an observed flux at 1.8 \micron\ of 15 nJy. 
At this wavelength,
the imaging sensitivity of an 8m NGST is 1 nJy ($10^4$ s exposure and
$10\sigma$ detection threshold), while the moderate resolution ($\lambda/\Delta
\lambda=1000$) spectroscopic limit is about 50 times higher ($10^5$ s exposure
per resolution element and $10\sigma$ detection threshold) (Stockman \etal
1997). The several weeks period of peak rest-frame blue luminosity would be
stretched by a factor of $(1+z)$ to few months. Figure 3 shows the cumulative
number of Type II events expected per year per $4'\times 4'$ field. Depending
on the history of star formation at high redshifts, the NGST should detect
between 7 (in the merging model) and 15 (in the monolithic collapse scenario)
Type II SNe per field per year in the interval $2<z<4$, therefore providing a
way to distinguish between different scenarios of galaxy formation. The
possibility of detecting Type II SNe at $z\gta5$ from an early population of
galaxies has been investigated by Miralda-Escud\'e \& Rees (1997). By assuming
these are responsible for the generation of all the metals observed in the
Lyman-$\alpha$ forest at high redshifts, a high baryon density
($\Omega_bh_{50}^2=0.1$), and an average metallicity of $0.01Z_\odot$,
Miralda-Escud\'e \& Rees estimate the NGST should observe about 16 SN II per
field per year with $z\gta 5$. Note, however, that a metallicity smaller by a
factor $\sim 10$ compared to the value adopted by these authors has been
recently derived by Songaila (1997). For comparison, the models discussed in
this talk predict between 1 and 10 Type II SNe per field per year with
$z\gta 4$. 

\end{itemize}

\end{document}